# ELECTRON CLOUD OBSERVATIONS AND PREDICTIONS AT KEKB, PEP-II AND SUPERB FACTORIES

H. Fukuma[#], KEK, Oho1-1, Tsukuba, Ibaraki 305-0801, Japan

## Abstract

Electron cloud observations at B factories, i.e. KEKB and PEP-II, are reviewed. Predictions of electron cloud effects at Super B factories, i.e. SuperB and Super KEKB, are also reviewed.

## KEKB

First, a history about the electron cloud effect in KEKB is briefly summarized, then some details are described.

A coupled bunch instability (CBI) caused by the electron cloud was known in the middle of the design stage of KEKB through the observation at KEK Photon Factory PF [1] and subsequent theoretical work by K. Ohmi [2]. If no cares were taken, a short growth time of the CBI of 0.4ms due to the electron cloud was predicted at KEKB. A solenoid winding was proposed as a measure against CBI in the design report of KEKB [4]. The simulation suggested that the growth time is decreased to 5ms with the solenoid winding. The original design of a vacuum chamber which was a round copper chamber without anteroom was left unchanged. At the beginning of the operation of KEKB, a vertical beam size blowup was observed at the low energy positron ring (LER) [5]. Speculating that the blowup was caused by the electron cloud, solenoids were installed. At that time a model of the blowup caused by the electron cloud was proposed by K. Ohmi and F. Zimmermann [3]. The model explained the blowup as the head-tail instability induced by the electron cloud. The luminosity was increased as a result of the solenoid winding. Although the solenoids finally covered 95% of the drift space, the decrease of the specific luminosity was still observed in a three RF bucket spacing pattern and not resolved at the end of operation of KEKB.

The solenoid winding was the main mitigation method against the electron cloud. At the early stage of the winding, the solenoids covered 35% of the drift space. The strength of the solenoid field was about 45 Gauss [7]. A threshold current of the blowup and the luminosity were raised after applying the solenoid field [6]. Major installation of the solenoids has finished in early 2002. Finally, 95% of the drift space was covered by the solenoids as described above. Although the blowup was relaxed as the winding progressed, the solenoid winding was not enough to suppress the luminosity drop at three RF bucket spacing. Place where the remaining electron cloud stay was not identified.

Vertical sidebands, which are supposed to be an indication of the head-tail instability, were observed in the signal of the transverse dipole oscillation of bunches [10]. Sideband appeared at the upper side of a betatron tune. The beam current where the sidebands appeared was coincided with that where a drop of the specific luminosity began [11]. A simulation successfully reproduced the sidebands in the case that the cloud size is twenty times larger than the beam size [12].

At the end of the operation without crab cavities a fill pattern was straight 3.5 bucket spacing where three RF bucket and four RF bucket spacing were repeated alternately. The 3.27 bucket spacing, where the sequence of the bunch spacing was (333433343334334) in the unit of the bucket, was tried. The result showed that the specific luminosity in 3.27 bucket spacing was about 5 % lower than that in 3.5 bucket spacing [13]. The backward bunch between two bunches separated by four RF buckets had the highest specific luminosity. A single beam measurement showed that bunches which had the lower specific luminosity had higher betatron sideband peaks.

Mode spectrum of the CBI was strongly dependent on the existence of the solenoid field, which showed that the CBI was caused by the electron cloud. Observed mode spectra without the solenoid field showed that the horizontal and vertical spectra had very similar patterns, which were consistent with a simulation assuming that photoelectrons were produced uniformly around the chamber wall [15]. With the solenoid field, simulated mode spectrum assuming the solenoid field of 10 Gauss explained the observation although 45 Gauss was actually applied. This may indicate that electrons stay nearer to the beam position than expected. A low mode frequency peak appeared in both the simulation and the measurement. The peak was explained by the rotation frequency of the spiral trajectory of electrons along the chamber surface [16]. Measured maximum vertical growth time was 0.4ms which was consistent with a simulation assuming low secondary emission yield (SEY) of 1.0 [14].

Betatron tune shift gives an estimate of the electron density [9]. In the case of solenoid-off, the horizontal and vertical tunes increased along the bunch train then saturated [17]. The electron distribution seemed round because the horizontal and vertical tune shifts were almost the same, which was consistent with the observation of the CBI spectra. The estimated electron density from the tune shift at saturation was $1.1 \times 10^{12} m^{-3}$ which was roughly consistent with simulations [8]. In the case of solenoid-on, the horizontal tune shift was reduced to almost zero while the vertical tune shift was reduced only by 40 %. The reason of the different effect of the solenoids on the horizontal and vertical tune shifts is not understood yet. This may imply the distribution of the electron cloud flattens with solenoid-on.

A method to measure the electron cloud density near beam was proposed and applied to KEKB [18]. Applying a bias voltage of about 1 kV to a grid of a retarding field analyzer (RFA) selects near-beam electrons which have relatively high energy due to a strong kick by the beam. The method was useful for estimating the electron cloud density for the development of vacuum components to reduce the electron cloud. The method was applied to the measurement of the electron cloud density in a quadrupole and a solenoid as well as the drift space [19]. The result showed that the electron density was $3 \times 10^{10}$ m$^{-3}$ in the quadrupole and $5 \times 10^{7}$ m$^{-3}$ in the solenoid of 50 Gauss.

---

[#]hitoshi.fukuma@kek.jp

## PEP-II

First, a history about the electron cloud effect in PEP-II is briefly summarized, then some details are described.

The CBI caused by the electron cloud was studied by M. Furman and G. Lambertson at the design stage of PEP-II [22]. According to their studies, arc chambers which were ante-chambers made of aluminum were coated by TiN to reduce the secondary electrons. At the beginning of the operation of PEP-II, non-linear pressure rise was observed in straight sections where chambers were round and non-coated stainless steel chambers [25]. The pressure rise was accompanied by the beam size blowup. Droop on the luminosity along the train was also observed [24]. The solenoid winding and the optimization of the fill pattern with mini-gaps were main measures against the electron cloud [25, 26] as well as TiN coating of the arc chambers. After completion of the solenoid winding in almost the whole ring, the mini-gaps in by-2 pattern (i.e. two RF bucket spacing) were gradually reduced and eventually eliminated. Making a large tune space by machine tuning might help eliminate the mini-gaps [21]. PEP-II finally achieved a straight by-2 pattern at 3.2A. In the end, the electron cloud instability did not degrade the peak luminosity [20].

The electron cloud effect in PEP-II was studied by a specially developed simulation code POSINST [22]. The code has a sophisticated model of secondary electron emission [23], takes into account of the elliptical cross section of a chamber and the space charge force by the electron cloud. The code predicted the average density of the electron cloud of $\sim 4.1 \times 10^{11}$ electrons/m$^3$ at the pumping straight section and $\sim 1.1 \times 10^{12}$ electrons/m$^3$ at the bend section [22]. Since the length of the pumping straight and the bend were 7.15m and 0.45m respectively, the pumping straight made major contribution. The growth time of the CBI was in the range of 1–2 ms by the simulation.

In early commissioning period of the low energy ring (LER), a sharp nonlinear pressure increase was observed at straight sections as described before. Increase of pump current came mostly from multipacting electrons entering the pump from the beam chamber [25]. The electron multipacting was detected in all drift sections of the LER straights independent of the level of synchrotron radiation.

The solenoid winding was applied at PEP-II [25, 26] as KEKB. The last section in the straights was energized Mar. 2001. The installation of the solenoids in the LER arcs was finished by Feb. 2002. Maximum solenoid field was about 30 Gauss. In the summer of 2003, several improvements were done. Additional 50 Gauss solenoids were installed on drift chambers in all straight sections. The field was doubled at pumping Ts, transitions and small drift chambers for all LER straight sections. Additional solenoids were installed for "no sextupole" girders in all LER arcs. Major solenoid windings finished in 2003.

It was understood that the straight sections of the ring were the dominant parts that had electron cloud effects. Luminosity increased by about 25% when 120m straight solenoid sections were energized [25]. Switching solenoids "on" and "off" changed the beam sizes only with high beam current. The beam size changed in both horizontal and vertical planes.

To see the effect of the arc solenoids, Arc 7A solenoids were turned off and on. The electron current was measured by an electrode installed in a chamber. Without the solenoid field, the electrode current increased not exponentially but linearly with a threshold in a TiN coated arc vacuum chamber. Turning off solenoids over 100 m did not degrade the luminosity.

Fill patterns were extensively optimized in PEP-II [29]. The bunch spacing was reduced and the mini-gaps were gradually decreased and finally eliminated to achieve a straight by-2 pattern. The progress to improve the fill patterns can be seen in several reports at EPAC and PAC [28].

Measurement of the single beam size blowup was studied by a gated camera [27]. Effectiveness of the mini-gaps to reduce the blowup was confirmed. Beam size growth was evident in the horizontal plane. On the other hand the blowup always occurred in the vertical plane in the simulation [30]. The reason of this discrepancy is not explained yet.

Reported data of the CBI in PEP-II are few. According to ref. [31], measured growth rate is consistent with simulations. Mode spectrum does not seem to agree with simulations.

A novel measurement technique named microwave transmission measurement [33] was applied at PEP-II [32]. The electron cloud (i.e. electron plasma) affects the propagation of the electromagnetic (EM) wave. The resulting phase shift of the wave is proportional to the electron density. The train gap modulates the electron density at the revolution frequency. The modulation appears as sidebands of the EM carrier whose peaks are proportional to the electron density. The effectiveness of the solenoid field was measured by this method [32]. The measurement showed that 10 Gauss of the solenoid field was enough to confine the electron cloud near the beam pipe walls.

## SUPERB

The latest information on the study of the electron cloud in SuperB is given by T. Demma in this workshop.

The results obtained before the workshop are shortly summarized here [34, 35].

The positron beam is stored in the high energy ring (HER) of 6.7 GeV in SuperB. The beam current of HER is 1.9 A, the number of bunches 978, the bunch spacing 1.2 m and the synchrotron tune 0.01. A threshold of the single bunch instability was estimated by a strong-strong code CMAD. Instability threshold was $\rho_e = 4 \times 10^{11} \text{m}^{-3}$.

The SuperB typically do not have long field free regions. For the most part of the ring the beam pipe is surrounded by magnets. Fraction where the dipoles cover the ring is 50%. The electron density in arc bend regions was evaluated by ECLOUD. The number of primary electrons was adjusted in order to take into account of the reduction of the electron yield by the ante-chamber. The result shows that if 99% of the photons are absorbed by the ante-chamber and SEY is 1.2, the electron density is below the threshold of the single bunch instability.

Clearing electrodes are considered as one of the mitigation methods. Simulations show that a bias voltage of 1 kV is sufficient to suppress the electron cloud formation [35].

## SUPERKEKB

The positron beam is stored in the low energy ring (LER) of 4 GeV in SuperKEKB. The beam current of LER is 3.6 A, the number of bunches 2500, the bunch spacing 1.2m and the synchrotron tune 0.026. A threshold electron density of the single bunch instability estimated by a stability condition is $2.7 \times 10^{11}$ m$^{-3}$ [36]. A threshold electron density calculated by a simulation [37] is $2.2 \times 10^{11}$ m$^{-3}$, which is consistent with the analytic estimate.

Growth time of the CBI was estimated at the threshold electron density of the single bunch instability [37]. The obtained growth time was 50 turns. The CBI could be damped by the bunch feedback system. Therefore the single bunch and the coupled bunch instability will be suppressed if the electron density is less than $2.2 \times 10^{11}$ m$^{-3}$. Thus the target electron density near beam against the electron cloud instabilities was taken to be less than $1 \times 10^{11}$ m$^{-3}$.

The electron density near beam in SuperKEKB was estimated to be $5 \times 10^{12}$ m$^{-3}$ based on results from measurements done at KEKB assuming a round copper pipe with a diameter of 94 mm, no solenoid field, 4 ns bunch spacing and 1 mA/bunch [38]. The cloud densities in a drift space, a bending magnet and a wiggler magnet were $8 \times 10^{12}$ m$^{-3}$, $1 \times 10^{12}$ m$^{-3}$ and $4 \times 10^{12}$ m$^{-3}$, respectively. Main contribution comes from the drift space. Integrated cloud density along the orbit in the drift space takes up about 80% of total integrated cloud density.

Various mitigation techniques were compared based on studies at KEK [38, 41]. Taking the electron density in a copper chamber as a standard, the reduction factor of the electron density is 1/50 with a 50 Gauss solenoid, 1/5 by an ante-chamber, 3/5 by TiN coating, 1/10 by grooves on top and bottom in a bend and 1/100 with a clearing electrode in a bend.

Following measures are to be applied at the SuperKEKB LER [39, 41]. In the drift space in arc sections aluminum ante-chambers with TiN coating and the solenoid winding are used. Bend chambers are TiN coated aluminum ante-chambers with grooved surface. Wiggler chambers are copper ante-chambers with clearing electrodes. Taking these measures, the electron density near beam is expected to be less than $1.0 \times 10^{11}$ m$^{-3}$.

A simulation showed that a threshold electron density of the single bunch instability at the positron damping ring of SuperKEKB was near the estimated electron density if no cares were taken [40]. While the integrated electron density by the simulation was $1.35 \times 10^{14}$ m$^{-2}$, the threshold of the single bunch instability by an analytic estimation was $1.57 \times 10^{14}$ m$^{-2}$. In order to keep enough margins against the electron cloud instability, TiN coating was considered as well as increasing the synchrotron tune from 0.004 to 0.015. Increase of the synchrotron is necessary for mitigating CSR instability in any case. A simulation showed that the integrated electron density was reduced to $0.51 \times 10^{14}$ m$^{-2}$ with TIN coating assuming SEY of one. The estimated threshold electron density of the single bunch instability in the optics with the raised synchrotron tune was $1.1 \times 10^{13}$ m$^{-3}$ which corresponded to the integrated electron density of $15 \times 10^{14}$ m$^{-2}$. The electron density is well below the threshold. Measures to be taken against the electron cloud at the damping ring are, aluminum ante-chambers with TiN coating, grooved surface on the wall of dipole chambers and solenoids at straight sections. The ante-chambers are required in order to install photon masks in any case.

## ACKNOWLEDGMENTS

The author would like to thank T. Demma, K. Ohmi, M. Pivi and L. Wang for informing him with several materials referred in this paper. He would like to express his special thanks to M. Sullivan for providing him information of PEP-II operation.

## REFERENCES

[1] M. Izawa, Y. Sato and T. Toyomasu, Phys. Rev. Lett. 74, 5044 (1995).
[2] K. Ohmi, Phys. Rev. Lett. 75, 1526 (1995).
[3] K. Ohmi and F. Zimmermann, Phys. Rev. Lett., 85, 3821(2000).
[4] KEKB Design Report, KEK report 95-7.
[5] H. Fukuma et al., EPAC'00, 1122 (2000).
[6] H. Fukuma et al., HEAC2001, Tsukuba,March, 2001.
[7] H. Fukuma et al., AIP conference proceedings 642, 354 (2002).
[8] F. Zimmermann, CERN-SL-2000-017 (2000).
[9] K. Ohmi et al., APAC'01, 445 (2001).
[10] J.W. Flanagan et al., Phys. Rev. Lett. 94, 054801 (2005).

[11] J.W. Flanagan et al., KEK Proceedings 2007-10, 15 (2007).
[12] E. Benedetto et al., PAC07, 4033 (2007).
[13] Y. Funakoshi et al., EPAC'06, 610 (2006).
[14] M. Tobiyama et al., PRST-AB PRST-AB 9, 012801 (2006).
[15] S. S. Win et al., ECLOUD'02, 199 (2002).
[16] S. S. Win et al., PRST-AB 8, 094401 (2005).
[17] T. Ieiri et al., EPAC06, 2101 (2006).
[18] K. Kanazawa et al., PAC'05, 1054 (2005).
[19] K. Kanazawa and H. Fukuma, IPAC'10, 2015 (2010).
[20] J. T. Seeman et al., EPAC'08, 946 (2008).
[21] M. Sullivan, private communications.
[22] M. A. Furman and G. R. Lambertson, EPAC'96, M. A. Furman and G. R. Lambertson, PAC'97, M. A. Furman and G. R. Lambertson, KEK Proceedings 97-17, 170 (1997).
[23] M. A. Furman and M. Pivi, PRST-AB 5, 124404 (2002).
[24] A. Kulikov et al., PAC'01, 1903 (2001).
[25] A. Kulikov et al., ECLOUD'04, 21 (2005).
[26] U. Wienands, the 13th ICFA Beam Dynamics Mini-Workshop, Upton, December 2003.
[27] R. Holtzapple, SLAC-PUB-9222 (2002).
[28] A. Kulikov et al., PAC'01, 1903 (2001). J. Seeman et al., PAC'03, 2297 (2003). J. Seeman et al., EPAC'04, 875 (2004).
[29] F. J. Decker et al., EPAC'00, 403 (2000), F. J. Decker et al., PAC'01, 1963 (2001), F. J. Decker, ECLOUD'02, 2002, F. J. Decker et al., EPAC'02, 392 (2002), F. J. Decker et al., EPAC'04, 842 (2004).
[30] Y. Cai, ECLOUD'02, 141 (2004).
[31] R. Akre et al., talk given at PEP-II MAC, 2006.
[32] S. De Santis et al., Phys. Rev. Lett., 100 (2008).
[33] T. Kroyer et al., ECLOUD'04, 89 (2004).
[34] T. Demma, IPAC'10, 2012 (2010).
[35] SuperB Progress Reports (2010).
[36] K. Ohmi, talk given at the seminar in Fermilab, June 14, 2011.
[37] Y. Susaki and K. Ohmi, IPAC'10, 1545 (2010).
[38] Y. Suetsugu, a talk given at the 15th KEKB Accelerator Review Committee (2010).
[39] Y. Suetsugu, a talk given at the 16th KEKB Accelerator Review Committee (2011).
[40] M. Kikuchi et al., IPAC'10, 1641 (2010).
[41] K. Shibata, in this proceedings.